# Towards commands recommender system in BIM authoring tool using Transformers

Changyu Du, Zihan Deng, Stavros Nousias, André Borrmann
Technical University of Munich, Germany
changyu.du@tum.de

**Abstract.** The complexity of BIM software presents significant barriers to the widespread adoption of BIM and model-based design within the Architecture, Engineering, and Construction (AEC) sector. End-users frequently express concerns regarding the additional effort required to create a sufficiently detailed BIM model when compared with conventional 2D drafting. This study explores the potential of sequential recommendation systems to accelerate the BIM modeling process. By treating BIM software commands as recommendable items, we introduce a novel end-to-end approach that predicts the next-best command based on user historical interactions. Our framework extensively preprocesses real-world, large-scale BIM log data, utilizes the transformer architectures from the latest large language models as the backbone network, and ultimately results in a prototype that provides real-time command suggestions within the BIM authoring tool Vectorworks. Subsequent experiments validated that our proposed model outperforms the previous study, demonstrating the immense potential of the recommendation system in enhancing design efficiency.

## 1. Introduction

The widespread adoption of BIM and model-based design in the AEC sector has been slowed by the fact that many users consider using BIM authoring tools to be an extra effort compared to traditional drawing-based design. The modern BIM software characteristically encompasses various aspects, disciplines, and systems of facility design. Nonetheless, such advancements have increased the complexity of interaction interfaces, making them much more expensive for users to learn. Due to the regularity and repetition exhibited by designers in the modeling process and the convoluted nature of command flows within the BIM design software, developing a decision-making predictive system that generates recommendations for the next-best actions could significantly reduce the time and errors during BIM modeling.

Sequential recommendation systems are essentially algorithms that predict user preferences by modeling the users' historical interaction sequences and actively proposing items that align with the users' interests. Such systems are widely used in online services like e-commerce and social media, especially in scenarios where users need to navigate through extensive collections of items and expect customized suggestions. In fact, designers face the same challenges as online service users when dealing with professional BIM authoring software due to the vast array of commands offered for various workflows and project requirements. While software vendors continuously work to streamline and enhance their UI for greater efficiency, designers still face the challenge of navigating through an extensive range of software commands to find the ones that meet their specific needs. This often requires significant software training and trial-and-error, even for users with extensive design experience, as they need to manually translate their knowledge into appropriate command flows for the design software.

Therefore, due to the similarity of the problems both aim to address, we treat commands in the BIM authoring tool as recommendable items, predicting the next-best commands through the user interaction sequences recorded in BIM log files. Utilizing the state-of-the-art transformer backbones originally designed for large language models (LLMs), we enhance the open-source deep recommender framework, Transformer4Rec (Moreira *et al.*, 2021), and propose an end-to-end command prediction pipeline. This led to the implementation of a prototype that enables real-time next-command recommendations for the BIM authoring tool Vectorworks. We conducted profound preprocessing of real-world, large-scale BIM log data, and performed training and extensive experiments on the resulting dataset, demonstrating the high potential of intelligent predictive systems in enhancing BIM design efficiency.

## 2. Related Works

### 2.1 BIM authoring command prediction

Sequences are suitable for representing successive user operations due to their intuitive nature. Pan et al. extracted command sequences from Revit journal files and grouped them into 14 classes that attempted to summarize the generic intent of the different commands at a high level. Eventually, an LSTM was employed to predict the class labels of upcoming commands (Pan and Zhang, 2020). However, their approach is restricted to predicting limited command categories rather than individual commands. Gao et al. developed a custom log in Rhino to combine modeling commands with their resulting 3D models, proposing the command-object graph to represent the modeling design behavior (Gao *et al.*, 2021). Their subsequent research (Gao *et al.*, 2022) extracted paths from the graphs to compose extensive command sequences, and used the native Transformer model to achieve the command-level prediction. Although the custom log can extract command sequences more effectively by integrating specific information of model elements compared to native log, its limitations lie in limited public access and the necessity for manual updates (Jang *et al.*, 2023). This results in a constrained dataset that may not accurately reflect real-world software usage. Furthermore, their basic transformer model does not fully leverage the potential side information in the log files.

### 2.2 Transformers for sequential recommendation

Recent developments in LLMs have showcased their incredible ability to understand language sequences. The underlying transformer architecture, originally designed for Natural Language Processing (NLP) tasks, has been widely adopted in other fields due to its scalability and ability to model complex dynamic patterns in sequence data efficiently (Gao *et al.*, 2022). In fact, the sequential processing of user interactions is similar to language modeling tasks. Many critical recommendation system architectures are inspired by and adapted from NLP, such as BERT4rec (Sun *et al.*, 2019). The success of the transformer architecture primarily lies in its proposed attention mechanism, which captures dependencies between representation pairs regardless of their distance in the sequence (Sun *et al.*, 2019). This allows it to model the dynamic and evolving nature of user sequential behavior effectively — much like the dynamic design process where designers' interests and steps may vary across different workflows and projects.

Transformer-based recommendation systems often incorporate side information as additional features to enrich the representation of user interaction sequences. This might include features like the user profile, product descriptions and images (Rashed et al., 2022). This approach differs from the transformers in the NLP field. In language modeling scenarios, there is usually no additional side information available. Transformer-based language models typically start by using a tokenizer to split the raw text into tokens and convert them into index encodings. The transformer architecture then learns the latent representations of the index sequences. Finally, depending on the NLP task, the prediction head can be designed for sentiment analysis, next token generation, etc. Transformer4rec (Moreira *et al.*, 2021) proposed a flexible framework that bridges the gap between NLP and sequential recommendation. Their work adapted the transformer architecture of language models for recommendation systems by replacing the tokenizer and the NLP head with an input module capable of accepting potential side information and a prediction head designed explicitly for recommendation tasks. They compared transformer backbones from various language models like BERT (Devlin *et al.*, 2019), XLNet (Yang *et al.*, 2019) in the task of generating recommendations for news and e-commerce scenarios, demonstrating state-of-the-art performance.

## 2.3 Research gaps

In summary, we identify the existing research gaps in the literature as follows:

- From the data perspective, current studies lack fine-grained command prediction based on large-scale native BIM log data gathered from real-world BIM tool usage.

- From the algorithm perspective, most research relies on basic sequence prediction models and has not explored the application of more advanced deep sequential recommendation systems, which have been highly successful in other fields. Additionally, considering the powerful capability of state-of-the-art LLMs in understanding and learning language sequences, integrating their model architectures into the command recommendation system would be a meaningful direction to explore.

- From the system integration perspective, current studies have not proposed an end-to-end pipeline that seamlessly integrates the prediction model into BIM authoring software for real-time command recommendation.

## 3. Methodology

We propose an end-to-end pipeline for predicting users' next actions in BIM authoring tools by conceptualizing commands as recommendable items utilizing anonymized log data collected through Vectorworks. The overall research framework is shown in Figure 1. We began by analyzing raw data, and proposed a detailed data filtering and preprocessing method to capture users' true software usage logic. Our model is built upon the Transformer4rec framework (Moreira *et al.*, 2021), utilizing transformer architectures from the state-of-the-art LLMs as backbone networks, and employs a parameter-efficient fine-tuning method for training. We evaluated the performance of different backbones, ultimately proposing a software architecture that deploys the best-performing model in real-time command recommendation scenarios.

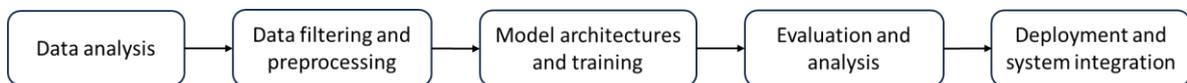

Figure 1 The overall research framework

## 3.1 Data analysis

The research employs a day's log data from 1000 anonymized users worldwide collected through Vectorworks, which contains 25,397,138 lines of records and 7 languages. A snippet of raw data content is shown in Figure 2. We cleaned the dataset by retaining only the essential components for further analysis: *session_anonymized* (anonymized modeling session ID), *timestamps* (when the command occurred), command *category*, and *message* (command content followed by a corresponding localization ID). Commands are categorized as *UNDO*, *Tool*, and *Menu*, with UNDO commands constituting nearly 95% of the data. An UNDO command represents an undoable/redoable event triggered in software and is the smallest unit used to record user actions. It includes various sub-actions for events, such as "End Event", "Event", "Undo Event", "Redo Event", and others. The term "Event" signifies the initiation of an event, whereas "End Event" denotes its successful completion, which we regard as sign of a valid action. "Tool" and "Menu" are higher-level commands, which are buttons that users can directly interact with on the UI. Our analysis reveals that higher-level commands either trigger an UNDO command (event) or do not lead to UNDO commands (events). Some events can be triggered directly by the cursor or keyboard without the higher-level buttons.

Figure 2 Content of the Vectorworks native log file

## 3.2 Data filtering and preprocessing

Figure 3 illustrates our proposed data filtering and preprocessing method. Based on the data analysis, we initially focused on removing irrelevant commands from semantic and statistical perspectives. This included records unrelated to user actions, such as software internal events; commands that frequently occur but hold little significance for BIM modeling and recommendations, like zooming and panning views; and aborted and unfinished commands.

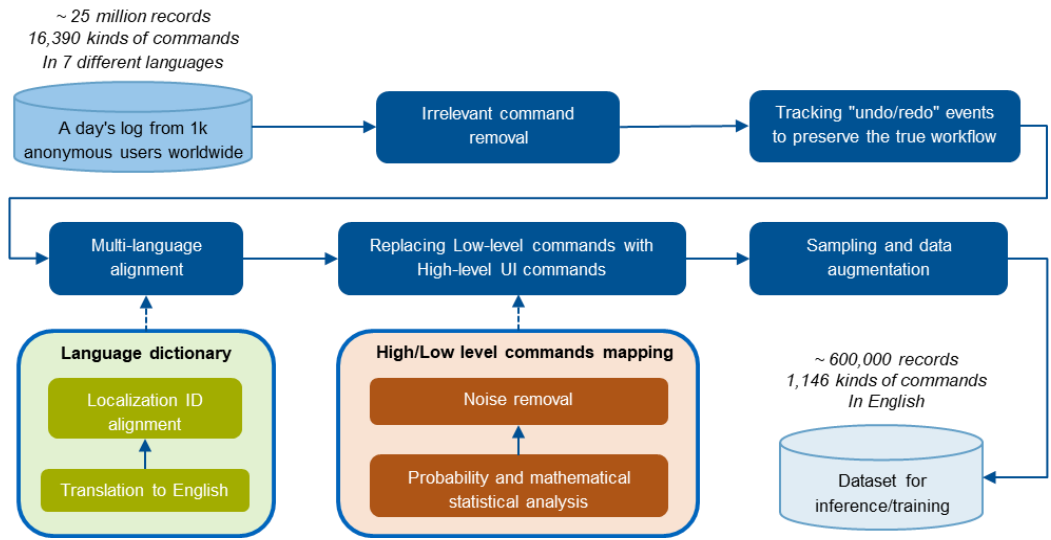

Figure 3 The proposed data filtering and preprocessing pipeline. After going through these steps, we successfully extracted 1146 kinds of commands from native logs, covering different workflows, projects, and disciplines.

In the subsequent step, logic is implemented to track the actual command flow in log records, including removing undo operations and any events they negate. Similarly, the corresponding events that were previously undone must be restored for the redo operations.

Given the global origin of the dataset, it encompasses a diverse array of languages, necessitating a rigorous multi-language alignment to avert potential errors from language-specific commands while aligning the software usage patterns of Vectorworks users worldwide with a uniform representation. We undertake a two-tier translation approach to ensure uniformity across command data. Initially, the Google Translate API is employed to automatically translate all commands into English. Subsequently, for enhanced accuracy, these translations are reviewed and manually refined. This manual alignment is guided by the corresponding localization ID, ensuring precise correspondence. Through this process, we construct a multilingual dictionary, enabling consistent multi-language alignment within the dataset.

We assume that the primary focus of command recommendations is on "Tool" and "Menu", which are interactable via UI buttons, as they are inherently more intuitive to the user. Therefore, substituting low-level events (e.g., *End Event: Clip Surface (169)*, *End Event: Create Database Worksheet (19)*) with the high-level "Tool" (e.g., *Tool: Clipping (-226)*) or "Menu" (e.g., *Menu: Create Report - (-172) (0)*) commands that trigger them is essential. As there is no documentation of mappings between low-level Events and high-level Tool/Menu within Vectorworks, we adopt a statistical method to establish the linkage between them and delineate causality. This involves cataloging the sequence in which actions occur, pinpointing the first finished event that follows or immediately follows a "Tool/Menu" operation. Through this method, we calculate the probability of events being consequent to "Tool/Menu" commands, resulting in a distinctive probability distribution pattern for these events. The uniform distribution suggests the "Tool/Menu" did not precipitate an event, while a pronounced distribution indicating a prominent event indicates a direct trigger by the "Tool/Menu" command. Subsequent to this probabilistic analysis, a manual review is conducted to validate the inferred relationships, ensuring high/low-level command mapping precision. Leveraging this mapping, we can substitute low-level events with their corresponding high-level "Tool/Menu" initiators and excise any unfinished "Tool/Menu" commands that do not instigate an event. This streamlines the dataset into a more concise and coherent form, accurately reflecting the authentic modeling process. Additionally, we remove any ambiguous commands deemed meaningless, usually due to error logging by third-party plugins.

We augment command sequence representation by integrating side information from logs, including command types, names, and time intervals. Noting the vast variance in log session lengths (0 to tens of thousands of records), we capped sequence length at 100 to balance context richness with computational efficiency, given transformer models' context window constraints. To mimic production scenarios, we randomly split extended sessions into 5-100 length subsequences, taking 80% of them as the training set and the rest for validation. Table 1 shows the statistics of the final dataset obtained after the data filtering and preprocessing pipeline.

Table 1 Statistics of the final dataset

| Session amount | Command classes | Command category | Record amount | Examples |
|---|---|---|---|---|
| 13,613 | 1,146 | UNDO | 431,020 | End Event: Delete (58), End Event: Drag (75) … |
| | | Tool | 103,227 | Tool: Move by Points (-352), Tool: Reshape (-355) … |
| | | Menu | 67,522 | Menu: Paste - (-29) (0), Menu: Duplicate - (-33) (0) … |

### 3.3 Model architecture

Figure 4 illustrates the proposed model architecture built upon the Transformer4rec (Moreira *et al.*, 2021). In the input sequence, each item (command) is not only represented by a unique command ID (analogous to the use of token ID to represent words and sub-words in NLP models), but also possesses features derived from various side information obtained by the data preprocessing pipeline, including command type, normalized time interval, and text embedding acquired by inputting the command name into a pre-trained sentence-transformer model (Reimers and Gurevych, 2019).

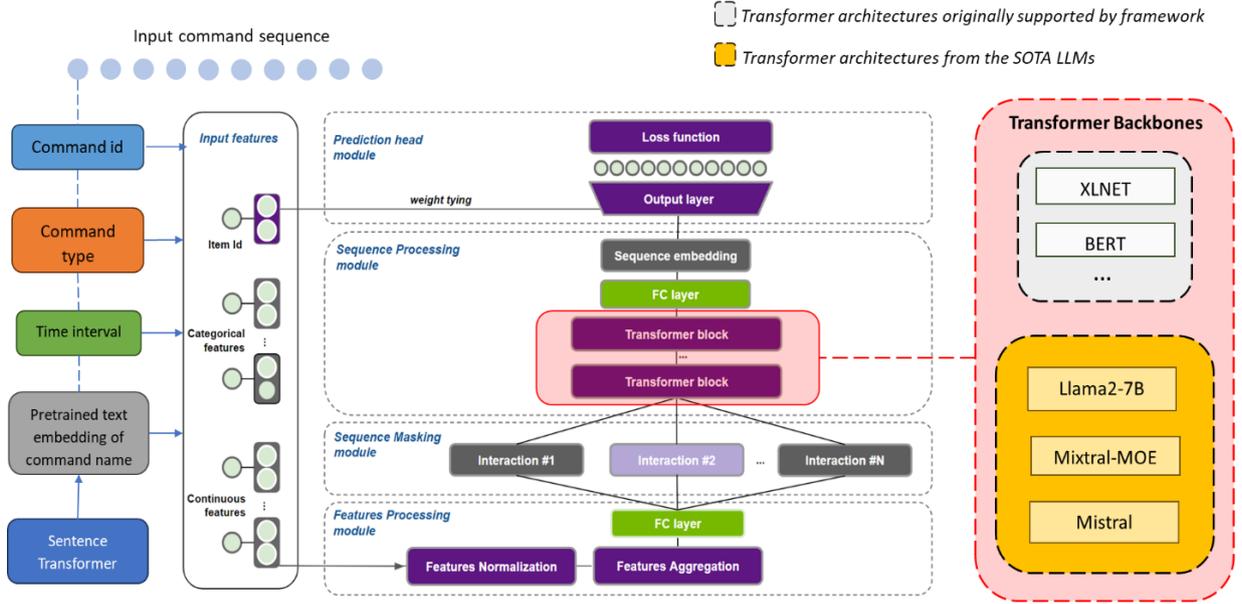

Figure 4 Proposed model architecture based on the Transformer4rec framework (Moreira et al., 2021). We expanded the original framework to support the transformer architectures from the latest LLMs.

Within the feature processing module, the command IDs are first converted into high-dimensional, learnable vectors through an embedding layer. Afterward, the item (command) embeddings, normalized continuous features (time interval), categorical features (command type), and the pre-trained text embeddings are projected into the same dimensional vector space through separate fully connected layers, then concatenated together, and finally transformed into the tensor dimensions accepted by the transformer modules through another fully connected layer.

Before feeding the input sequence into the transformer backbones, the sequence masking module masks the items to be predicted and saves the ground truth labels for the prediction head, enabling the model to be trained in a self-supervised manner. We mainly use two masking patterns in our study: Causal Language Modeling (CLM) and Masked Language Modeling (MLM). CLM ensures that the next item is predicted given the context of its previous items, so it masks the right context, allowing the model only to see information on the left — transformer models like GPT that are decoder-only use this mechanism. Conversely, MLM is more like filling in the blanks; it randomly masks 15% of the items in the sequence, and the model can access bidirectional context information to predict the masked items. Transformers based on BERT mostly use this mechanism. It is worth mentioning that since this approach leaks future information during training, we ensure that only the last item in the sequence is masked during inference to make it compatible with the next-item prediction task (Moreira *et al.*, 2021).

The sequence processing module contains stacked transformer blocks that learn the latent features of each item in the high-dimensional masked sequences. We expanded this part to support the transformer architectures from the latest open-source LLMs, including:

- **Llama2** (Touvron *et al.*, 2023) proposed by Meta, is one of the most popular decoder-only LLMs in the open-source community. It has made improvements to the native transformer architecture, such as using pre-normalization, i.e., normalizing the input of each transformer sub-layer, applying the SwiGLU activation function instead of ReLU, and replacing the absolute positional embeddings with rotary positional embeddings (RoPE).

- **Mistral** (Jiang *et al.*, 2023), is architecturally very similar to Llama, but utilizes techniques such as grouped-query attention (GQA) and sliding window attention (SWA) to speed up inference and process long sequences more efficiently.

- **Mixtral-MoE** (Jiang *et al.*, 2024), a Mixtrue of Experts (MoE) model that outperformed Llama2 and GPT-3.5 on most benchmarks. The concept of MoE aims to establish a regulatory mechanism for systems composed of multiple individual networks. In such systems, each network (expert) processes a different subset of training samples, focusing on a specific area of the input space. A gating network determines which experts handle the input by deciding the weights assigned to each expert and merging their outputs (Sanseviero *et al.*, 2023). The architecture of Mixtral is identical to Mistral's, except that they replace the feedforward network (FFN) within the transformer layers with MOE layers that consist of 8 feedforward blocks (i.e., experts).

Finally, the prediction head module uses an output layer followed by a softmax layer to predict the class label of the N+1$^{th}$ item based on the latent representations of the N$^{th}$ item in the output sequences of the previous module. The weight-tying technique lets the output layer share the weights with the item embedding layer, thus reducing the parameter amounts. The authors of Transformer4rec reported that bundled weighs can improve the model's prediction results (Moreira *et al.*, 2021). We replaced weight-tying with a linear layer mapping the output to the label dimensions and did not observe a significant difference in the prediction metrics in our dataset. Nonetheless, we still default to using this technique in our experiments to reduce the model parameter size. Last but not least, the cross entropy loss is used as the loss function.

### 3.4 Training strategy

Due to the large number of parameters in primitive architectures of these LLMs, which exceeded our hardware capabilities, we adapted the original architecture of the backbone models other than Llama2, e.g., instead of using a transformer with 32 or more layers, we reduced the number of parameters by using 2 or 4 layers. We trained such models from scratch using a regular self-supervised training method. Regarding the model with Llama2 backbone, we kept the 7B-parameters architectural setting of Llama2 and fine-tuned it using QLoRA (Dettmers *et al.*, 2023) based on the pre-trained weights. QLoRA (Quantized Low-Rank Adaptation) is a parameter-efficient fine-tuning method. It starts by quantizing and freezing the pre-trained model weights. Then, it injects trainable rank decomposition matrices into every layer of the transformer architecture, significantly reducing the number of trainable parameters. With this approach, we can fine-tune large models on a single GPU. Table 2 presents information on the various backbone networks utilized in our study. For comparison, we selected the best-performing transformer backbones reported in the Transformer4rec, including XLNet and BERT, along with their configurations.

Table 2 The information on different transformer backbones used in our study

| Transformer backbone | Total parameters | Trainable parameters | Architectural settings |
|---|---|---|---|
| Llama2 | 6,688,355,328 | 0.91% | 32 heads, 32 layers |
| Mixtral-MoE | 552,112,128 | 100% | 8 heads, 2 layers, 8 experts |
| Mistral | 630,632,448 | 100% | 8 heads, 4 layers |
| XLNET | 8,530,432 | 100% | 8 heads, 2 layers |
| BERT | 19,350,016 | 100% | 8 heads, 4 layers |

## 4. Results and analysis

All experiments were run on a workstation with a 24GB NVIDIA TITAN RTX GPU and 32 GB RAM. For XLNet and BERT, we employed the MLM masking approach, following the best-performing configurations reported in Transformer4rec. For other models, we utilized the CLM method due to their decoder-only architecture. Regardless of which masking approach is used, during evaluation, we ensure that only the last item in the input sequence is masked, allowing the model to make predictions for this item.

Table 3 shows the performance of our model on the validation set using different transformer backbones. We utilized Recall@K and NDCG@K as evaluation metrics to calculate the top K accuracy in the recommendation list containing all commands. NDCG (Normalized Discounted Cumulative Gain) considers the ranking position of the relevant command in the recommendation list, with higher rankings yielding higher values. Recall, on the other hand, only checks whether the relevant command is among the top K items. Since in a production environment, we aim to present users with the top few superior recommendations rather than the probability distribution for all commands, we consider that choosing K as 5 or 10 is reasonable.

Table 3 Evaluation metrics of the next-command prediction

| Transformer backbone | Recall@5 | NDCG@5 | Recall@10 | NDCG@10 |
|---|---|---|---|---|
| Llama2 | 77.15% | 63.92% | 85.94% | 66.80% |
| Mixtral-MoE | **78.10%** | **65.01%** | 86.16% | 67.65% |
| Mistral | 77.26% | 64.62% | **86.49%** | **67.66%** |
| XLNET | 74.16% | 61.76% | 83.50% | 64.82% |
| BERT | 68.83% | 52.30% | 81.45% | 56.43% |

The table shows that our models based on the latest LLM architectures (Llama2, Mixtral-MoE, Mistral) outperform the best architectures reported in the original Transformer4rec paper (XLNET, BERT) across all metrics. The performance of the three latest LLM backbones is relatively similar, with Mixtral-MoE slightly standing out due to its efficient and unique architectural design. It is noteworthy that despite Llama2 having a significantly larger number of parameters than the others, its performance on our dataset is comparable to that of smaller networks. We believe that the scale and complexity of our dataset may not fully leverage the capabilities of such large models. Additionally, due to hardware constraints, extra quantization operations and the insertion of low-rank matrices may lead to some loss in model accuracy compared to conventional fine-tuning methods.

## 5. Deployment and system integration

We developed a software prototype (Figure 5) to deploy trained models and recommended the next-best commands to users in real-time. Figure 6 shows the underlying software architecture. Considering the computational cost of inference, we deployed the model and feature engineering pipeline on Nvidia's Triton Inference Server hosted on a remote GPU server. The GPU server is connected to a local PC running Vectorworks via SSH, and data is transmitted using the GRPC/HTTP protocol. On the local PC, we implemented an application using Vue.js and FastAPI to poll the real-time log contents generated by Vectorworks. In the backend, we implemented the pipeline introduced in Section 3.2 to preprocess and filter data, send it to the inference server, and dynamically display the received predictions on the frontend.

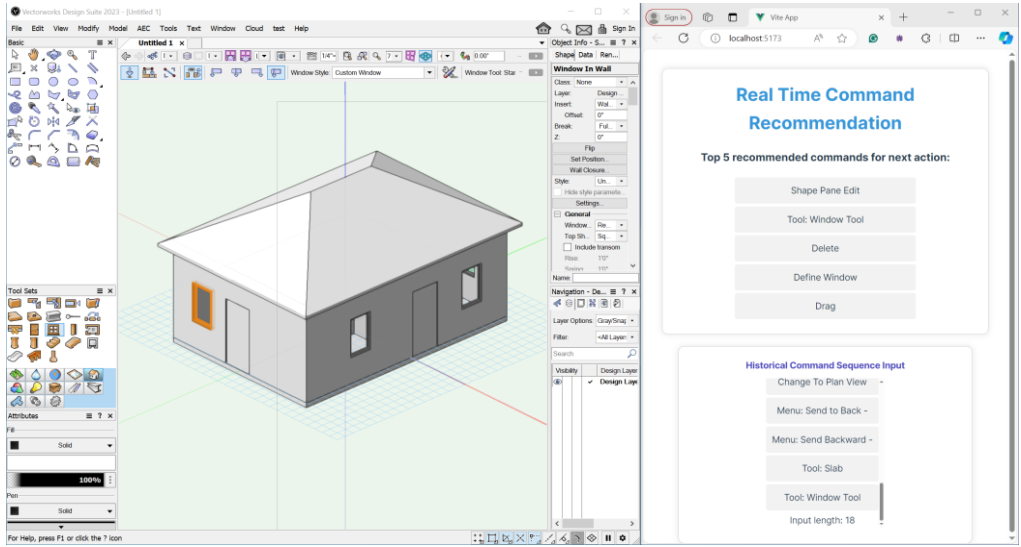

Figure 5 Software prototype running in parallel to Vectorworks, predicting next actions in real-time during the design.

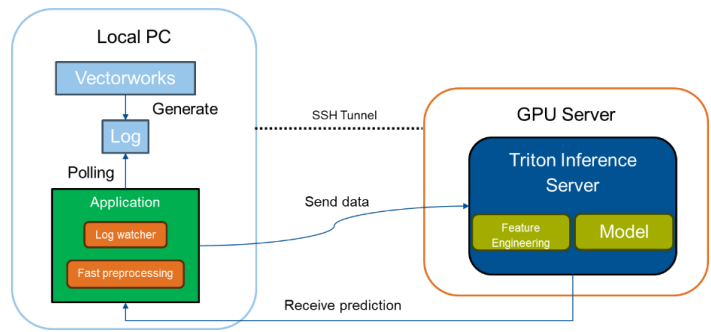

Figure 6 Proposed software architecture of deploying model for real-time command recommendations in Vectorworks. Our standalone application is not coupled with Vectorworks and can be combined with any other BIM authoring tool.

## 6. Conclusion and future works

In this paper, we introduce and implement an end-to-end real-time BIM command recommendation system based on the large-scale native log data from Vectorworks, filling the research gaps outlined in Section 2.3. Our proposed model leverages the transformer architectures of the latest LLMs as its backbone, and significantly outperforms the previous study on real-world BIM log datasets. Going forward, we plan to conduct more comprehensive ablation studies to explore the optimal methods for leveraging and aggregating side information from log files. Additionally, we aim to train our model on more extensive datasets using better hardware to fully harness the capabilities of large models, providing next-best action recommendations throughout the entire software usage lifecycle. Test users will be recruited to assess our recommendation system's impact on their workflows in real-world projects.

**Acknowledgment**

This work is funded by Nemetschek Group, which is gratefully acknowledged. We sincerely appreciate the data and licensing support provided by Vectorworks, Inc.

# References

Dettmers, T., Pagnoni, A., Holtzman, A. and Zettlemoyer, L. (2023) 'QLoRA: Efficient Finetuning of Quantized LLMs', in *Advances in Neural Information Processing Systems*, pp. 10088–10115. Available at: https://proceedings.neurips.cc/paper_files/paper/2023/file/1feb87871436031bdc0f2beaa62a049b-Paper-Conference.pdf.

Devlin, J., Chang, M.-W., Lee, K. and Toutanova, K. (2019) 'BERT: Pre-training of Deep Bidirectional Transformers for Language Understanding', in *Proceedings of the 2019 Conference of the North {A}merican Chapter of the Association for Computational Linguistics: Human Language Technologies, Volume 1 (Long and Short Papers)*. Association for Computational Linguistics, pp. 4171–4186. Available at: https://doi.org/10.18653/v1/N19-1423.

Gao, W., Wu, C., Huang, W., Lin, B. and Su, X. (2021) 'A data structure for studying 3D modeling design behavior based on event logs', *Automation in Construction*, 132. Available at: https://doi.org/10.1016/j.autcon.2021.103967.

Gao, W., Zhang, X., He, Q., Lin, B. and Huang, W. (2022) 'Command prediction based on early 3D modeling design logs by deep neural networks', *Automation in Construction*, 133. Available at: https://doi.org/10.1016/j.autcon.2021.104026.

Jang, S., Lee, G., Shin, S. and Roh, H. (2023) 'Lexicon-based content analysis of BIM logs for diverse BIM log mining use cases', *Advanced Engineering Informatics*, 57. Available at: https://doi.org/10.1016/j.aei.2023.102079.

Jiang, A.Q., Sablayrolles, A., Mensch, A., Bamford, C., Chaplot, D.S., Casas, D. de las, Bressand, F., Lengyel, G., Lample, G., Saulnier, L., Lavaud, L.R., Lachaux, M.-A., Stock, P., Scao, T. Le, Lavril, T., Wang, T., Lacroix, T. and Sayed, W. El (2023) 'Mistral 7B'. Available at: http://arxiv.org/abs/2310.06825.

Jiang, A.Q., Sablayrolles, A., Roux, A., Mensch, A., Savary, B., Bamford, C., Chaplot, D.S., Casas, D. de las, Hanna, E.B., Bressand, F., Lengyel, G., Bour, G., Lample, G., Lavaud, L.R., Saulnier, L., Lachaux, M.-A., Stock, P., Subramanian, S., Yang, S., Antoniak, S., Scao, T. Le, Gervet, T., Lavril, T., Wang, T., Lacroix, T. and Sayed, W. El (2024) 'Mixtral of Experts'. Available at: http://arxiv.org/abs/2401.04088.

Moreira, G.D.S.P., Rabhi, S., Lee, J.M., Ak, R. and Oldridge, E. (2021) 'Transformers4Rec: Bridging the Gap between NLP and sequential/session-based recommendation', in *RecSys 2021 - 15th ACM Conference on Recommender Systems*. Association for Computing Machinery, Inc, pp. 143–153. Available at: https://doi.org/10.1145/3460231.3474255.

Pan, Y. and Zhang, L. (2020) 'BIM log mining: Learning and predicting design commands', *Automation in Construction*, 112. Available at: https://doi.org/10.1016/j.autcon.2020.103107.

Rashed, A., Elsayed, S. and Schmidt-Thieme, L. (2022) 'Context and Attribute-Aware Sequential Recommendation via Cross-Attention', in *RecSys 2022 - Proceedings of the 16th ACM Conference on Recommender Systems*. Association for Computing Machinery, Inc, pp. 71–80. Available at: https://doi.org/10.1145/3523227.3546777.

Reimers, N. and Gurevych, I. (2019) 'Sentence-BERT: Sentence Embeddings using Siamese BERT-Networks'. Available at: http://arxiv.org/abs/1908.10084.

Sanseviero, O., Tunstall, L., Schmid, P., Mangrulkar, S., Belkada, Y. and Cuenca, P. (2023) *Mixture of Experts Explained*, *Hugging Face Blog*. Available at: https://huggingface.co/blog/moe (Accessed: 4 March 2024).

Sun, F., Liu, J., Wu, J., Pei, C., Lin, X., Ou, W. and Jiang, P. (2019) 'Bert4rec: Sequential recommendation with bidirectional encoder representations from transformer', in *International Conference on Information and Knowledge Management, Proceedings*. Association for Computing Machinery, pp. 1441–1450. Available at: https://doi.org/10.1145/3357384.3357895.

Touvron, H., Martin, L., Stone, K., Albert, P., Almahairi, A., Babaei, Y., Bashlykov, N., Batra, S., Bhargava, P., Bhosale, S., Bikel, D., Blecher, L., Ferrer, C.C., Chen, M., Cucurull, G., Esiobu, D., Fernandes, J., Fu, J., Fu, W., Fuller, B., Gao, C., Goswami, V., Goyal, N., Hartshorn, A., Hosseini, S., Hou, R., Inan, H., Kardas, M., Kerkez, V., Khabsa, M., Kloumann, I., Korenev, A., Koura, P.S., Lachaux, M.-A., Lavril, T., Lee, J., Liskovich, D., Lu, Y., Mao, Y., Martinet, X., Mihaylov, T., Mishra, P., Molybog, I., Nie, Y., Poulton, A., Reizenstein, J., Rungta, R., Saladi, K., Schelten, A., Silva, R., Smith, E.M., Subramanian, R., Tan, X.E., Tang, B., Taylor, R., Williams, A., Kuan, J.X., Xu, P., Yan, Z., Zarov, I., Zhang, Y., Fan, A., Kambadur, M., Narang, S., Rodriguez, A., Stojnic, R., Edunov, S. and Scialom, T. (2023) 'Llama 2: Open Foundation and Fine-Tuned Chat Models'. Available at: http://arxiv.org/abs/2307.09288.

Yang, Z., Dai, Z., Yang, Y., Carbonell, J., Salakhutdinov, R. and Le, Q. V (2019) 'XLNet: Generalized Autoregressive Pretraining for Language Understanding', in *Proceedings of the 33rd International Conference on Neural Information Processing Systems*. Red Hook, NY, USA: Curran Associates Inc.